\documentclass[10pt,twocolumn,twoside]{article}

\usepackage[utf8]{inputenc}
\usepackage[T1]{fontenc}
\usepackage[english]{babel} 
\usepackage{CJKutf8} % arXiv 必须使用 CJKutf8 处理中文

\usepackage{graphicx}
\usepackage{amsmath,amssymb,amsfonts}
\usepackage{float}
\usepackage{xcolor}
\usepackage{titlesec}
\usepackage{authblk} 
\usepackage{fancyhdr}
\usepackage{abstract}
\usepackage{tipa} % 保持你原来的 tipa 支持
\usepackage[numbers,sort&compress]{natbib}
\usepackage{orcidlink}
\usepackage{hyperref}

\definecolor{pnasblue}{RGB}{0, 114, 188}
\hypersetup{colorlinks=true, linkcolor=pnasblue, citecolor=pnasblue, urlcolor=pnasblue}

\usepackage[left=1.5cm,right=1.5cm,top=2cm,bottom=2cm]{geometry}

\titleformat{\section}{\large\bfseries\color{pnasblue}\sffamily}{\thesection}{1em}{}
\titleformat{\subsection}{\normalsize\bfseries\sffamily}{\thesubsection}{1em}{}

\title{\huge \bfseries \sffamily The Maintenance and Necessity of Universal Rules: Scale, Hierarchy, the Cost of Justice, and Civilizational Development \\ \Large \color{gray} \huge \textnormal{普世规则的维护与必然：规模、层级、正义成本与文明发展}}

\author[a,b,1]{Li Tuobang 李拓邦\orcidlink{0000-0002-2257-2603}}
\affil[a]{Independent Researcher, Zhaoqing, Guangdong, China 独立研究员，肇庆，广东，中国}
\affil[b]{University of California, Berkeley 加州大学伯克利分校}

\affil[1]{To whom correspondence should be addressed. 联系方式。E-mail: lituobang@hotmail.com}

\newcommand{\acknow}[1]{\section*{Acknowledgment 致谢} {\small #1}}

\begin{document}

% 启动中文环境 (gbsn 是简体宋体，gkai 是楷体)
\begin{CJK*}{UTF8}{gbsn}

\twocolumn[
  \begin{@twocolumnfalse}
    \maketitle
    \begin{abstract}
Building upon previous research, this paper further explores the topological foundations for maintaining universal rules within ultra-large-scale societies. It finds that in small-scale societies, absolute egalitarianism and the rule of law can be compatible through peer monitoring within a fully connected network. However, in ultra-large-scale societies, to maintain high-dimensional rules capable of protecting innovation and property rights, a complex hierarchical structure including "high-fragility" nodes must be constructed. Through quantitative analysis of power structures, this paper proves that a flattened, two-tier structure inevitably leads to the degradation of the rule of law. Only a social topology with sufficient hierarchical depth can escape the deathly trap of the Leviathan while expanding in scale, thereby sustaining the dynamic evolution of civilization.

本文在前期研究的基础上，进一步探讨了超大规模社会中普世规则维护的拓扑基础。本文发现，在小规模社会中，绝对平均与法治可通过全连接网络的同侪监督兼容；但在超大规模社会中，为了维持一种能够保护创新与产权的高维规则，必须构建一套包含“高脆弱性”节点的复杂层级结构。本文通过对权力结构的定量分析证明，扁平化的两层结构必然导致法治的退化，而只有具备足够层级深度的社会拓扑，才能在规模扩张的同时，逃逸利维坦的死寂陷阱，维持文明的动态演化。

    \end{abstract}
    
    % 如果需要在英文版脚注显示中文引用信息
    %\vspace{0.5cm}
   % {\small \color{gray} The Chinese Version of this article has been published in Finance and Management: \\
  %  \href{https://ijssrr.com/journal/article/view/3198}{https://ijssrr.com/journal/article/view/3198}}
  %  \vspace{1cm}
  \end{@twocolumnfalse}
]

\section*{Introduction 引言}
In traditional discussions of social justice, "absolute egalitarianism" is often envisioned as the ultimate ideal for eliminating predation and inequality. However, when viewed through the lens of evolutionary institutionalism, this extreme pursuit of individual uniformity is fundamentally a violation of systemic evolutionary logic. Within the governance blueprint of the Leviathan, absolute egalitarianism is not a ladder to a utopian society, but rather a tool for clearing the field to achieve monolithic control.

在关于社会公正的传统讨论中，“绝对平均主义”常被寄托为一种消除掠夺与不平等的终极理想。然而，若从演化制度学的视角审视，这种对个体一致性的极致追求，本质上是对系统演化逻辑的根本性违背。在利维坦（Leviathan）的治理蓝图中，绝对平均主义并非通向大同的阶梯，而是实现单极权力控制（Monolithic Control）的清场工具。

We have established in previous arguments that multilayered predatory structures eventually collapse under the weight of excessive bargaining costs\cite{Li_2026}. Meanwhile, the Monolithic Leviathan, though achieving a compressed atomization of society through enforced individual uniformity, freezes complex dissipative structures into a static grid. This leads civilization into a state of stasis and "heat death." Simultaneously, we have identified the Rule of Law as the sole path to transcending the "war of all against all," marking the transition from disordered predation to orderly prosperity\cite{Li_2026}.

我们在前期的论证中已经阐明，多层掠夺结构最终会因博弈成本过高而不可避免地走向坍塌\cite{Li_2026}。与此同时，单一利维坦（Monolithic Leviathan）虽然通过强制性的个体一致性实现了社会的原子化压缩，却将复杂的耗散结构冻结为静态网格，导致文明陷入死寂状态。与此同时，我们也确立了法治作为摆脱“所有人对所有人的战争”，从无序掠夺迈向秩序繁荣的唯一路径\cite{Li_2026}。

This paper aims to explore a core proposition: In ultra-large-scale societies, how do universal rules escape the stasis trap of the Leviathan to maintain dynamic evolution\cite{hobbes1651leviathan}? Classical political science has long focused on the decisive influence of population size on the nature of a polity. Montesquieu (1748), in The Spirit of the Laws, perceptively noted that the form of government must match the territorial scale, arguing that ultra-large states would inevitably slide into despotism if their structures remained unchanged\cite{de2005spirit}. Diamond (1997) further argued from an evolutionary perspective that in a tribe of several hundred people, everyone knows one another or is connected through kinship. If a dispute arises between two individuals, the surrounding kin will collectively mediate\cite{diamond1999guns}. However, in a society of tens of thousands, a dispute between two strangers can escalate into a chain reaction of murderous vendettas without a central authority to mediate. Consequently, as the population grows, the difficulty of collective decision-making rises exponentially. Large populations cannot reach consensus through universal consultation; therefore, power must be surrendered to a bureaucratic system or a centralized authority.

本文旨在探讨一个核心命题：在超大规模社会中，普世规则究竟如何逃逸利维坦的死寂陷阱而维持动态演化\cite{hobbes1651leviathan}？经典政治学长期关注人口规模对政体性质的决定性影响。孟德斯鸠 (Montesquieu, 1748) 在《论法的精神》中敏锐地指出，政体形式必须与领土尺度相匹配，认为超大规模国家若不改变结构，必然滑向专制\cite{de2005spirit}。戴蒙德 (Diamond, 1997) 进一步从演化视角论证，在一个几百人的部落中，每个人都互相认识，或者通过亲戚关系互相联系\cite{diamond1999guns}。如果两个人发生了争执，由于周围的人都是他们的亲戚，大家会合力劝解。但在一个成千上万人的社会里，两个陌生人相遇发生争执时，如果没有一个中央机构来调停，这种冲突将演变成连锁反应式的仇杀。因此随着人数增加，群体决策的难度呈指数级上升。大规模人群无法通过全员协商达成共识，因此必须将权力让渡给一个科层体系或中央集权。

Regarding the restraint of power, Weber (1922) proposed two types of authority: "Charismatic Authority," which relies on the leader's charm, and "Legal-Rational Authority," which depends on procedures and offices\cite{weber1978economy}. Tocqueville (1835) and Arendt (1951) both emphasized the defensive function of "intermediate associations" against atomized tyranny\cite{de1835government,arendt1973origins}. Arendt profoundly noted that an atomized mass is the breeding ground for totalitarianism\cite{arendt1973origins}. In New Institutional Economics, Coase (1937) and Williamson (1975) view hierarchy as an alternative solution for reducing market transaction costs\cite{coase1937economica,williamson1975markets}. This paper extends this logic to the maintenance of the Rule of Law: the existence of hierarchy is fundamentally necessitated by the fact that the cost of maintaining rules undergoes a quadratic explosion in ultra-large-scale societies. As a high-dimensional public good, the stability of the Rule of Law is directly proportional to the hierarchical depth of the system. Through a quantitative analysis of power structures, this paper demonstrates that only a social topology with sufficient hierarchical depth can escape the stasis trap of the Leviathan while expanding in scale. Ultimately, the developmental zenith of a civilization is determined by the complexity of the legal levers its structural framework can sustain.

关于权力如何被制约，韦伯 (Weber, 1922) 提出了两种权威类型。其中，“克里斯玛型权威”依赖领袖魅力，而“法理型权威”依赖程序与职位\cite{weber1978economy}。托克维尔 (Tocqueville, 1835) 与 阿伦特 (Arendt, 1951) 均强调了中间层级（Intermediate Associations）在对抗原子化暴政中的防御功能\cite{de1835government,arendt1973origins}。阿伦特深刻指出，原子化大众正是极权主义繁衍的温床\cite{arendt1973origins}。科斯 (Coase, 1937) 与 威廉姆森 (Williamson, 1975) 在新制度经济学中将层级视为降低市场交易费用的替代方案\cite{coase1937economica,williamson1975markets}。本文将此逻辑扩展至法治维护：层级的存在，本质上是由于规则维护成本在超大规模社会中面临“平方级爆炸”的物理约束。法治作为一种高维公共产品，其系统的稳态水平与层级拓扑的深度互为正相关。通过对权力结构的定量分析，本文证明：唯有具备充足层级深度的社会拓扑，方能在规模扩张下，逃逸利维坦式（Leviathan）的死寂陷阱。文明的发展能级，最终取决于该系统所能负载的法治杠杆之复杂度。

\section*{The Scale Limit of Absolute Egalitarianism 绝对平均主义的规模限制} 

To analyze the long-term viability of absolute egalitarianism, we establish a set of basic principles derived from institutional economics, social physics, and law:

为了分析绝对平均主义的长期生存能力，我们确立了一套源自制度经济学，社会物理学和法学的基本原则：

\begin{enumerate}\item \textbf{Principle of Heterogeneity 多样性原则:} As systemic entropy increases, the variance among participants expands. No two participants possess identical cost structures or biological resilience.

随着系统熵的增加，参与者之间的差异性也会扩大。没有任何两个参与者拥有完全相同的成本结构或生理韧性。\item \textbf{Principle of Rational Self-Interest 理性自私原则:} While altruism exists, the presence of self-interested actors is a statistical inevitability due to heterogeneity.

尽管利他主义存在，但由于多样性的存在，自私行为者的出现是统计学上的必然。\item \textbf{Principle of Limited Totality 总量有限原则:} Output and matter are bounded by the laws of thermodynamics; no system can externalize costs indefinitely.

产出和物质受热力学定律的约束；没有任何系统可以无限期地将成本外部化。\item \textbf{Selection Bias in Competition 竞争中的选择偏好:} Entities occupying dominant positions tend to expand their proportional share within the system over time, while less efficient entities diminish.

随着时间的推移，占据优势地位的实体往往会在体系中扩大其比例份额，而效率低下的实体则会减少。\item \textbf{Principle of Differential Association 差别接触原则:}The successful operational modes of dominant entities are mimicked by subordinate entities, a process defined as "Differential Association"\cite{sutherland1947differential}. 

优势实体的成功运作模式会被处于劣势的实体所模仿，这一过程被定义为“差异接触” \cite{sutherland1947differential}。\item \textbf{Law of Large Numbers 大数原则:} As the population scale and observation time increase, these principles converge into deterministic social trends.

随着人口规模和观察时间的增加，这些原则会收敛为确定性的社会趋势。\end{enumerate}
When a complex system adopts "absolute egalitarianism" as its distribution rule, it essentially activates a "negative feedback" mechanism.

当一个复杂系统采用“绝对平均主义”作为分配规则时，系统实质上开启了一个“负向反馈”机制。

Due to the coupling of the Principle of Heterogeneity and the Principle of Rational Self-Interest, "arbitrageurs" inevitably emerge within the system. Absolute egalitarianism entails a total decoupling of contribution from reward, providing institutional collateral for "free-riding." Under the Law of Large Numbers, as long as the expected utility of this strategy (i.e., the average share) exceeds its game-theoretic cost, this behavioral pattern will rapidly propagate through the Principle of Differential Association \cite{sutherland1947differential}. Ultimately, governed by the Principle of Limited Totality, the system is destined for extinction.

由于多样性原则与理性自私原则的耦合，系统内不可避免地出现“套利者”。绝对平均主义意味着贡献与收益的完全脱钩，这为“不劳而获”提供了制度性保障。在大数原则的作用下，只要该策略的预期收益（即平均分配所得）大于其博弈成本，该行为模式就会通过差异接触原则迅速在系统内扩散 \cite{sutherland1947differential}。最终，受限于总量有限原则，该系统必然走向消亡。

If the system attempts to sustain its egalitarian form by establishing a punitive apparatus against free-riding, it falls into two irreconcilable logical traps:

如果该体系试图维持其绝对平均的形态，针对不劳而获建立一套惩罚体系，它将陷入以下两个不可调和的逻辑陷阱：

Trap I: The Paradox of Inequality Induced by Punishment

陷阱一：惩罚导致的不平等悖论

The establishment of a punitive mechanism to penalize "free-riders" through the confiscation of resources directly violates the definition of absolute equality. As the scale of arbitrageurs expands, such punishment leads to the emergence of a massive "underclass." At this juncture, the system slides instantaneously from "absolute equality" into "extreme inequality," causing governance costs to surge exponentially.

若建立惩罚机制，通过剥夺物资来惩戒“不劳而获者”，则直接破坏了绝对平均的定义。随着套利者规模的扩大，这种惩罚会导致一个庞大的下层贱民群体（Underclass）的出现。此时，系统从“绝对平均”瞬间滑向了“极端不平等”，其治理成本将呈指数级增长。

Trap II: Management Entropy and "Overseer Costs"

陷阱二：管理熵与“监工成本”

Punishment is not cost-free. According to Dunbar (1992), the size of the human neocortex constrains the upper limit of stable social relationships to approximately 150 individuals \cite{dunbar1992neocortex}. Within this threshold, a system can rely on minimal information costs and peer pressure to suppress "free-riding" behavior. However, once the social scale $N$ crosses this biological constraint, the system must establish specialized punitive institutions to replace failed peer monitoring, leading to an exponential surge in management entropy. To maintain absolute egalitarianism, the system must sustain a professional group dedicated to strict punishment. This punitive group itself becomes a dominant entity within the system; due to the Principle of Limited Totality, to sustain the operation of this group, the system must further extract surplus value from bottom-level producers. This is a fatal evolutionary constraint. Suppose there exists another non-egalitarian system that naturally eliminates the incentive for free-riding through reward mechanisms (such as incentives for high producers). In long-term evolutionary competition, an egalitarian society burdened by the dual costs of "arbitrageurs" and the "punitive group" will exhibit significantly lower systemic efficiency compared to the former.

惩罚并非免费。根据 Dunbar (1992) 的研究，人类新皮层的大小限制了其能够维持稳定社会关系的上限（约 150 人）\cite{dunbar1992neocortex}。在这一阈值内，系统可以依靠极低的信息成本和熟人压力来抑制‘不劳而获’行为。然而，当社会规模 $N$ 跨越这一生物学约束，系统必须通过建立专门的惩罚机构来替代失效的同侪监督，从而导致了管理熵的指数级激增。要想维持绝对平均主义，系统必须供养一个专门负责严厉惩罚的专业团体。该惩罚团体本身成为系统的优势实体，由于总量有限原则，为了维持这一团体的运作，系统必须进一步压榨底层生产者的剩余价值。这是一个致命的演化约束。假设存在另一个非平均主义体系，其通过奖励机制（如对高产出者的激励）自然消弭了不劳而获的动力。在长期的演化竞争中，背负着“套利者”与“惩罚团体”双重成本负担的平均主义社会，其系统效率将远低于前者。

Thus, setting aside the fact that economic activities are usually not simple mining but require a high degree of subjective initiative—even in the case of mining, high rewards must be given to those who find ore. Otherwise, a problem arises: while finding ore may indeed be due to luck, if the reward for the person who finds it is the same as for those who do not, who would strive to mine? Who would expect to find anything? If a group of overseers is hired to supervise these miners, it often leads to numerous vicious incidents and widespread public resentment. It would be far more effective to give the money spent on overseers to the miners who actually find the ore. A rational mechanism design can make all participants happy, resulting in fewer conflicts and higher output. In contrast, an irrational mechanism usually leads to universal outrage with little to no effect.

所以，且不说经济活动通常不是单纯的挖矿，是需要高度主观能动性的。哪怕是挖矿，也是需要给挖到矿的人高额奖励的，否则就会产生一个问题，挖到矿的人固然是因为运气，但如果其得到的奖励，和没有挖到矿的人是一样的，那么谁又会去努力挖矿？谁又会期待可以挖到矿？如果请一帮监工去监督这些矿工挖矿，那常会产生很多恶性事件，搞得大家怨声载道，还不如把这个请监工的钱给挖到矿的矿工，效果还更好。合理的机制设定，可以让参与者都很开心，矛盾很少，产出也很高。而不合理的机制通常搞的天怒人怒。又没什么效果。

The global absolute equalization movements of the mid-20th century testified to the severity of this logic. Whether in the commune experiments in the U.S., the early Israeli Kibbutzim, or the absolute equalization movements in Soviet regions, these organizations eventually mostly moved toward extinction. Survivors followed only two evolutionary paths:

20世纪中叶的全球绝对平均化运动证明了这一逻辑的严酷。无论是在美国的公社试验、以色列的早期基布兹（Kibbutz），还是苏联地区的绝对平均化运动，这些组织最终大多走向消亡。幸存者仅剩下两条演化路径：

Alienation: Introducing hierarchy and punishment, effectively abandoning absolute egalitarianism, which usually evolves into a monolithic Leviathan structure. A typical case is Soviet Union.

异化：引入层级与惩罚，事实上放弃了绝对平均主义，最终通常会演化为单一利维坦结构。典型案例是苏联。

Micro-scaling: This strategy involves strictly limiting social scale to within the threshold allowed by Dunbar’s Number, utilizing the ultra-low monitoring costs of a "society of acquaintances" to hedge against the risks of free-riding. This scale is generally situated between 100 and 500 individuals. The experience of the Israeli Kibbutz validates the reliability of these figures. Through a comparative study of religious and secular kibbutzim, Sosis (2000) discovered that community scale is inversely proportional to the level of cooperation among members \cite{sosis2000religion}. At this scale, the social topology approximates a Fully Connected Graph, where the cost of information transmission $C_{info} \approx 0$. Violations are broadcast instantaneously, and the society can be maintained through "peer monitoring" without the need for additional hierarchical tiers. In contrast, large-scale kibbutzim are more prone to free-riding behaviors; consequently, after facing financial crises in the late 20th century, they transitioned from flat structures toward multi-tiered hierarchical structures \cite{abramitzky2008limits}.

微型化：将规模严格控制在邓巴数允许的范围以内，利用熟人社会的超低监督成本来对冲不劳而获的风险。这个规模大致在100-500人左右。以色列基布兹的经验也印证了这一数字的可靠性。Sosis (2000)通过对宗教性与世俗性基布兹的对比研究发现，社区规模（Scale）与成员间的合作水平呈反比\cite{sosis2000religion}。此规模下，社会拓扑接近全连接图（Fully Connected Graph）。信息传递成本 $C_{info} \approx 0$。违规行为会被瞬间广播，社会依靠“同侪监督”即可维持，无需额外层级。而大规模基布兹更容易出现搭便车行为，所以在 20 世纪末面临财务危机后，也从扁平化结构向多层级结构转型\cite{abramitzky2008limits}。

As Mao Zedong noted: "The source of absolute egalitarianism, like extreme democratization in politics, is a product of handicraft and small-peasant economy."\cite{mao1961selected} Small-peasant economy should be respected as a value as long as they respect the rule of law in the society, but a distinction must be made: absolute egalitarianism under a small-peasant economy is a "self-governing steady state" based on the scale of Dunbar’s number, whereas absolute egalitarianism under a monolithic Leviathan is an attempt to forcibly impose a deathly order upon an ultra-large-scale society.

正如毛泽东所指出的：“绝对平均主义来源，和政治上的极端民主化一样是手工业和小农经济的产物”\cite{mao1961selected}。小农经济作为一种价值观，只要他们尊重社会的法治规则，就应当予以尊重，但必须区分：小农经济下的绝对平均主义是基于邓巴数规模的“自治稳态”，而单一利维坦下的绝对平均主义则是试图在超大规模社会中强行推行一种死寂秩序。

\section*{Technical Maintenance of the Rule of Law and the Scaling Effect of Hierarchical Complexity 法治的技术性维护与层级复杂度的尺度效应}

As previously argued, in ultra-large-scale societies, multilayered predatory structures serve as the incubator for the Rule of Law, whereas a monolithic Leviathan structure precludes its developmen\cite{Li_2026}. To transition into a steady-state Rule of Law society, the focus must shift from the absolute equality of distribution to the absolute consistency of rules. Hayek (1960) posited that the essence of the Rule of Law lies in the universality and predictability of rules rather than the equality of outcomes \cite{hayek2020constitution}. Beyond the acceptance of a rigorous and universal legal framework, such a transition requires a corresponding maintenance system.

在前文中，我们已经论述过，在超大规模社会中，多层掠夺结构是法治的孵化器，而在单一利维坦结构下是无法发展出法治的\cite{Li_2026}。要想走入稳态的法治社会，必须从分配的绝对一致转向规则的绝对一致。哈耶克 (Hayek, 1960) 曾指出，法治的关键在于规则的普适性与预见性，而非结果的平等 \cite{hayek2020constitution}。而从多层掠夺结构走向法治社会，除了接受一套具有严密性与极高普适性的法治原则，还需要有相应的维护体系。

The Rule of Law is essentially a set of entropy-reversing universal rules. From a physics perspective, this implies that the Rule of Law is a high-energy dissipative structure \cite{prigogine1977self}. In ultra-large-scale societies, the cost of maintaining these rules explodes exponentially, transforming the technical maintenance of the Rule of Law into a steady-state equilibrium problem of computational complexity and graph topology.

法治（Rule of Law）本质上是一套逆熵的普世规则。在物理学视角下，这意味着法治是一个高能耗的耗散结构 (Dissipative Structure) \cite{prigogine1977self}。在超大规模社会中，维护规则的成本将呈指数级爆炸，法治的维护遂演变为一个计算复杂度与图论拓扑的稳态均衡问题。

We define the Cost of Justice Index ($J$):

我们定义法治维护成本指数($J$)：$$J= \frac{Power_{violator}}{\beta \epsilon}$$

Where $Power_{violator}$ represents the magnitude of the violator's power, and $\beta \in (0, 1]$ is the power fragility coefficient. This coefficient defines the degree of ease with which a power node is removed by the system when it violates universal rules. $\epsilon$ denotes the violation degree, where $\epsilon \in (0, 1]$. $\epsilon \to 1$ indicates a felony or explicit violation. For the sake of simplicity in the following discussion, we primarily focus on the case where $\epsilon = 1$.

其中，$Power_{violator}$ 是违规者的权力大小，$\beta \in (0, 1]$ 是权力脆弱性系数。该系数定义了当一个权力节点违背普世规则时，系统将其剔除的难易程度。$\epsilon$是违规程度。$\epsilon \in (0, 1]$。$\epsilon \to 1$：重罪。在下文讨论中，为简便讨论，主要讨论$\epsilon=1$的情况。

Low-fragility nodes ($\beta < 0.5$) typically correspond to populist politicians, or "Charismatic Authority." The power of such politicians stems directly from the fanatic support of the masses \cite{weber1978economy}. This support is characterized by extreme viscosity and irrationality. Even when the politician violates the Rule of Law, their supporters often refuse to accept punishment based on identity politics. To maintain the Rule of Law, other nodes must pay an immense price (such as launching a larger-scale mobilization) to provide a counterbalance.

低脆弱性节点（$\beta<0.5$）通常为民粹政客，即“克里斯玛型权威”，这类政客的权力直接来源于底层的狂热支持\cite{weber1978economy}。这些支持具有极高的粘性和非理性。即便政客违反法治，其支持者往往基于身份认同而拒绝接受惩罚。要维护法治，其他节点必须付出极大的代价（如发动更大规模的动员）才能制衡。

Suppose the system needs to counterbalance a politician with 1 million supporters and a power fragility coefficient of $0.1$. The "justice energy" required for mobilization would be:

假设要制衡拥有 $100$ 万支持者的权力脆弱性系数为0.1的政客，系统需要调动的“正义能量”为：$$J = \frac{1,000,000}{0.1} = 10,000,000 \text{ (equivalent units of supporters)}$$

This means either a coalition of 10 peers, each with 1 million supporters, or 33 medium-sized nodes, each with 300,000 supporters, is required.

这意味着要 $10$ 个拥有 $100$ 万支持者的同等权力者联合，或者 $33$ 个拥有 $30$ 万支持者的中型节点联合。

High-fragility nodes ($\beta > 0.5$) typically correspond to technocrats, or "Legal-rational Authority." Their power is derived solely from procedural authorization and professional office \cite{weber1978economy}. Once their behavior deviates from universal rules, their legitimacy collapses instantly, and the system can replace them or overturn their decisions at a very low cost.

高脆弱性节点（$\beta > 0.5$）通常为技术官僚，即“法理型权威”，其权力仅来源于程序授权和专业职位\cite{weber1978economy}。一旦其行为偏离普世规则，其合法性即刻瓦解，系统可以用极低的成本将其替换或推翻其决定。

Suppose a judge $J$ also possesses an influence covering $1$ million people, but their power originates from professional codes of conduct. Their fragility $\beta$ is extremely high (e.g., $\beta = 0.9$). Once their behavior deviates from universal rules, the "justice energy" the system needs to mobilize is:

假设一名法官 $J$ 的影响力同样覆盖 $100$ 万人，但其权力来自职业准则。其脆弱性 $\beta$ 极高（例如 $\beta = 0.9$），一旦其行为偏离普世规则，系统需要调动的“正义能量”为：$$J = \frac{1,000,000}{0.9} \approx 1,111,111$$

Counterbalancing them can be achieved with only a slight amount of system redundancy (for example, 4 nodes each possessing 300,000 supporters).

制衡他只需要稍微多一点点的系统冗余即可实现（例如 4 个拥有 30 万支持者的节点）。

After establishing the $J$ model, we face a more profound question: Why not directly counterbalance large nodes through an infinite number of micro-nodes? Having established the energy requirements, we must introduce Coordination Cost ($C_{coord}$). The total burden of a counterbalancing action and the Cost of Justice Index are closely related to coordination costs. Furthermore, coordination costs follow the formula for the number of edges in a complete graph.

在确立了 $J$ 模型后，我们面临一个更深刻的问题：为什么不直接通过无数个微型节点来制衡大节点？在确立了能量需求后，我们必须引入协调成本（Coordination Cost, $C_{coord}$）。制衡行动的总负担和法治维护成本指数与协调成本密切相关。而协调成本遵循全连接图的边数公式。

Assume that any collaboration must reach a consensus among all participating nodes. According to graph theory, the communication paths (coordination cost) required for $k$ nodes to achieve a complete connection is:

假设任何协作都必须在所有参与节点间达成一致。根据图论，$k$ 个节点实现全连接所需的通讯路径（协调成本）为：$$C_{coord}(k) = \frac{k(k-1)}{2} \approx O(k^2)$$

This implies that if one attempts to counterbalance a single large node by increasing the number of participants, the coordination cost will undergo a quadratic explosion. To counterbalance a violator, the system must aggregate $k$ justice nodes.

这意味着，如果试图通过增加参与者数量来制衡单个大节点，协调成本会出现平方级爆炸。为了制衡违规者，系统必须聚合 $k$ 个正义节点。

We compare the coordination overhead of two strategies: In the counterbalancing of a low-fragility node, it is necessary to mobilize 10 nodes each possessing 1 million supporters. Its coordination cost is $C_{coord} = \frac{10 \times 9}{2} = \mathbf{45}$. Meanwhile, the $J$ value of these 10 nodes is also significant, making them highly susceptible to evolving into uncontrolled new predators.

我们对比两种策略的协调开销：在低脆弱性节点的制衡中，需要调动 10 个各拥有 100 万支持者的节点。其协调成本$C_{coord} = \frac{10 \times 9}{2} = \mathbf{45}$。同时这10 个节点的$J$也不小，极易演化为不受控的新掠夺者。

To mitigate this risk, one might instead use 33 nodes each possessing 300,000 supporters. Its coordination cost is $C_{coord} = \frac{33 \times 32}{2} = \mathbf{528}$. Therefore, to reduce the $J$ value of the enforcement nodes, the system pays 11.7 times the coordination overhead.

为了降低风险，改用 33 个各拥有 30 万支持者的节点。其协调成本$C_{coord} = \frac{33 \times 32}{2} = \mathbf{528}$。所以为了降低执法节点的$J$，系统支付了 11.7 倍的协调开销。

In contrast, in the counterbalancing of a high-fragility node, only 4 nodes each possessing 300,000 supporters need to be mobilized. Its coordination cost is $C_{coord} = \frac{4 \times 3}{2} = \mathbf{6}$. Moreover, the $J$ value of these nodes is low, making the possibility of them evolving into uncontrolled new predators extremely unlikely.

而在高脆弱性节点的制衡中，需要调动 4 个各拥有 30 万支持者的节点。其协调成本$C_{coord} = \frac{4 \times 3}{2} = \mathbf{6}$。同时这6 个节点的$J$也不高，之后演化为不受控的新掠夺者的可能性极低。

Simultaneously, to satisfy the Coordination Cost Consistency Constraint and prevent $C_{coord}$ from exploding, the coordination scale $k$ at any single tier must be locked within a range of computational constants. As the total social population (or total number of nodes) $N$ increases, the only mathematical solution is to increase the hierarchical depth $L$. The relationship between the system's hierarchical depth $L$ and $N$ follows:

同时为了满足协调成本一致性约束，不让$C_{coord}$ 爆炸：即任何一层的协调规模 $k$ 必须锁定在一个计算常数范围。在社会总人口（或节点总数） $N$ 增加时，唯一的数学解就是增加层级深度 $L$。系统的层级深度 $L$ 与 $N$ 的关系遵循：$$N = k^L \implies L = \log_k N$$

This means that to keep coordination overhead within a safe "steady-state interval" affordable by the system, the hierarchical depth $L$ must grow logarithmically with $N$.

这意味着，为了让协调开销保持在一个系统可支付的安全的“稳态区间”，层级深度 $L$ 必须随 $N$ 呈对数级增长。

We define the system's Total Enforcement Cost ($C_{total}$) as the sum of the coordination costs of each tier. The Leviathan mode, by eliminating middle tiers through flattening, causes the cost of counterbalancing Leviathan to rapidly exceed the limits of what the system can pay:

我们将系统的总执法成本 ($C_{total}$) 定义为每一层协调成本的加总。利维坦模式通过扁平化消灭中层，这导致制衡利维坦的成本迅速突破系统所能支付的限度：$$C_{total} = C_{coord}(N) = \frac{N(N-1)}{2} \approx O(N^2)$$

However, in a multi-tiered structure that conforms to $L = \log_k N$, assuming the local coordination cost $C_{coord}(k)$ of each tier is a constant $\gamma$:
而在符合$L = \log_k N$的多层级结构中，假设每一层的局部协调成本 $C_{coord}(k)$ 为常数：$\gamma$。$$C_{total} = L \cdot \gamma = (\log_k N) \cdot \gamma \approx O(\log N)$$

By paying the depth cost of $L$ tiers, the system reduces total overhead from a quadratic level to a logarithmic level. Thus, the realization of a Rule of Law society becomes possible.

系统通过支付 $L$ 层的深度成本，将总开销从平方级降为了对数级。因此实现法治社会变成了可能。

However, the increase in hierarchical depth $L$ also introduces two secondary issues:

然而，层级深度 $L$ 的增加也带来了两个次生问题：

1. Information Loss: The transmission of information from top-level rules to terminal execution requires crossing $L$ nodes. If the transmission loss per tier is $\delta$, the global error will accumulate with $L$. If $L$ is too small, the system hits the $O(k^2)$ "Coordination Wall" (the Leviathan Trap). If $L$ is too large, the system hits the "Entropy Wall" of information transmission (the Bureaucratic Trap) \cite{merton1940bureaucratic}.

1，信息损耗：信息从顶层规则传导至末端执行，需要跨越 $L$ 个节点。如果每一层的传导损耗为 $\delta$，则全局误差会随 $L$ 累积。如果 $L$ 太小，系统会撞上 $O(k^2)$ 的协调墙（利维坦陷阱）。如果 $L$ 太大，系统会撞上信息传导的熵增墙（官僚主义陷阱）\cite{merton1940bureaucratic}。

2. Narrative Legitimacy: High-tier nodes must possess sufficient indicators to prove their superiority over lower-tier nodes in professional or ethical dimensions; otherwise, their legitimacy as maintainers of the Rule of Law will face challenges.

2，叙事合理性：高层节点必须具备足够的指标证明其在专业或伦理上优于低层节点，否则其作为法治维护者的合法性将面临挑战。

For democratically elected politicians, information loss and narrative legitimacy are not significant issues, as the scale of their support is determined entirely by the design of the electoral system. Elected politicians are indispensable; they are responsible for reflecting public demands and representing value orientations. However, elected politicians are typically low-fragility nodes, prone to evolving into uncontrolled new predators. Consequently, in the institutional designs of Western countries, elected politicians are restrained by checks and balances, while highly specialized, high-fragility nodes are the indispensable force for realizing a Rule of Law society, minimizing the local coordination cost $\gamma$ at each tier. Yet, such nodes are not available on demand. Once hierarchical depth reaches a certain threshold, the society may lack a sufficient supply of high-fragility nodes with professional indicators strong enough to prove their superiority over lower tiers. As a result, the system cannot sustain such a deep $L$ and, consequently, fails to maintain universal rules at a larger scale $N$.

对于民主选举的政客来说，信息损耗和叙事合理性都不是什么问题，因为其支持者的规模完全取决于选举制度的设定。民选政客是不可或缺的，他们负责反映民众诉求并代表价值取向。但民主选举的政客通常是低脆弱性节点，容易演化为不受控的新掠夺者。所以在欧美国家的制度设计中，民主选举的政客都被分权制衡，而高度专业化、高脆弱性的节点是实现法治社会不可或缺的力量，它们使每层的局部协调成本 $\gamma$ 极小化。可是这些节点并不是想有就有的。一旦层级深度达到比较高的程度，可能整个社会也没有多少高脆弱性节点有足够的指标能在专业性上证明自己确实在某些方面比低层的节点要优秀，所以系统就无法负载这么深的 $L$，进而无法在更大规模的 $N$ 上维持普世规则。

This explains why many developing nations, despite having democratic forms with checks and balances, can only maintain the Rule of Law regarding extremely severe violations. Due to their large population scales $N$, a lack of Rule of Law consciousness among voters, and politicians possessing extremely low fragility coefficients—combined with a lack of what Weber termed "Legal-rational Authority" \cite{weber1978economy}—the comprehensive cost of enforcement remains prohibitive. Furthermore, in practice, coordination costs are closely related with the time. The longer the coordination time, the higher the difficulty of reaching consensus. Thus, quadratic growth ($O(k^2)$) actually underestimates the true coordination cost.

这也解释了为何许多发展中国家虽有民主形式，却仅能在极严重的违规上维持法治。由于人口规模 $N$ 庞大，选民法治意识薄弱，政客往往具有极低的脆弱性系数，且缺乏韦伯所言的“法理型权威” \cite{weber1978economy}，导致综合执法成本依然高昂。此外，在实践中协调成本还与时间密切相关。协调时间越长，达成共识的难度越高。因此，平方级增长（$O(k^2)$）实际上还低估了真实的协调成本。

A multi-tiered structure does more than just reduce coordination costs via $L = \log_k N$; more importantly, it enables high-resolution governance. Lower tiers (such as local chambers of commerce or trade associations) handle small-scale, high-precision (low $\epsilon$) violations. Only when the scale of violation expands to affect higher dimensions is it reported upward through the tiers.

多层级结构不仅通过 $L = \log_k N$ 降低了协调成本，更重要的是，它实现了高分辨率治理。底层级（如地方商会、行业协会）负责处理极小规模、极高精细度（低 $\epsilon$）的违规。只有当违规规模扩大、影响到更高维度时，才会逐级上报。

An effective Rule of Law system is a "distributed computing steady-state." By increasing functional hierarchical depth and rationally allocating low-fragility and high-fragility nodes, it achieves the incremental leverage of "justice energy." Hierarchy exists not for predation, but to bring the cost of maintaining justice in ultra-large-scale societies within an affordable range through "dimensional reduction." Only when the cost of justice is sufficiently low can the Rule of Law be maintained at a relatively small scale (small $\epsilon$ of violation), thereby supporting activities that fundamentally rely on legal protection for extreme rule precision such as complex finance, high-end R\&D, and global division.

真正有效的法治系统是一个“分布式计算稳态”。它通过增加功能性层级深度，合理分配低脆弱性节点和高脆弱性节点，实现了对正义能量的逐级杠杆放大。层级不是为了掠夺，而是为了通过“维数降阶”，将超大规模社会的正义维护成本控制在可支付范围内。正义成本足够低才能在相对小的尺度上也能维护法治（违规程度较小），从而可以负载复杂的金融、高端研发和全球分工，因为这些活动依赖于对极小违规精度的法律保护。

\section*{Criminological Perspective 犯罪学视角}

In the preceding sections, we correlated the cost of justice maintenance $J$ with the degree of violation $\epsilon$. From a criminological perspective, the generation of $\epsilon$ is not a stochastic fluctuation but a product of the interplay between individual control factors and environmental incentives. Criminology generally recognizes three primary causes of crime, which are highly correlated with node fragility $\beta$ and the social hierarchical structure in our $J$ model:

在前文中，我们将法治维护成本 $J$ 关联于违规程度 $\epsilon$。从犯罪学（Criminology）视角看，$\epsilon$ 的生成并非随机涨落，而是个体控制因素与环境激励博弈的产物。犯罪学公认犯罪主因可归纳为三个维度，这与本文 $J$ 模型中的社会层级结构高度相关：

1. Internal Control Imbalance: Individuals with Narcissistic Personality Disorder (NPD) or Antisocial Personality Disorder (ASPD) exhibit neuroanatomical deficits, such as reduced gray matter in the anterior insula or abnormalities in the prefrontal cortex, which can be detected via MRI \cite{raine2000reduced,schulze2013gray,schulze2014structural}. These organic lesions lead to a "functional loss" in emotional regulation and empathy, predisposing these nodes to extreme non-cooperative strategies and a high subjective propensity for violation. However, these individuals constitute a small percentage of the general population.

1. 内部控制失调： 自恋型人格障碍（NPD）或反社会型人格障碍（ASPD）患者在神经解剖学上表现出前岛叶灰质缺乏或前额叶皮层功能异常，这些病变可以通过核磁共振（MRI）检出 \cite{raine2000reduced,schulze2013gray,schulze2014structural}。这种器质性病变导致节点在情绪调节与同理心上出现“功能性缺失”，使其在博弈中倾向于选择极端的非合作策略，呈现出极高的主观违规倾向。但在社会总体人口中，这类个体的占比极低。

2. External Control Failure (Deterioration of the Rule-of-Law Environment): As described by the Broken Windows Theory, when minor illegal acts in a community go unpunished, the authority of the system's "meta-rules" is eroded \cite{kelling1982broken}. In the $J$ model, this implies an increased environmental tolerance for violations, causing external controls to fail and inducing potential offenders to cross the action threshold. A society with a robust hierarchical structure can maintain the Rule of Law, fostering a positive feedback loop for a healthy environment. Conversely, a society lacking the necessary hierarchy for law enforcement is prone to inducing crime. Social Control Theory further posits that swift and certain punishment increases the cost of violation, thereby suppressing $\epsilon$ \cite{hirschi2017causes}. This essentially forces a node's decision back toward a Nash equilibrium by raising the expected value of punishment.

2. 外部控制失效（法治环境败坏）： 正如“破窗效应”（Broken Windows Theory）所述，当一个社区的轻微违法行为未被惩罚，系统的“元规则”权威便会发生侵蚀 \cite{kelling1982broken}。在 $J$ 模型中，这意味着环境对违规的容忍度上升，导致外部控制因素（External Controls）失效，诱发潜在犯罪分子跨越行动阈值。一个拥有良好层级结构的社会能够更有效地维护法治，从而形成正向反馈的健康环境；反之，若社会缺乏维护法治所必需的层级深度，则极易诱发犯罪。与之对应，社会控制理论（Social Control Theory）认为，迅速而严厉的惩罚可以提高违规成本，从而压制 $\epsilon$ \cite{hirschi2017causes}。这在本质上是通过提升惩罚的期望值，强制将节点的博弈决策拉回到纳什均衡点。

3. Memetic Learning and Differential Association: Differential Association Theory posits that criminal behavior is learned. If an individual is exposed to more "pro-delinquent definitions" than "anti-delinquent" ones, their decision-making algorithm shifts \cite{sutherland1992principles}. Research indicates that social learning theory (a socio-psychological version of differential association) can explain up to 68\% of the variance in criminal behavior \cite{akers2017social}. This proves that the attributes of "neighboring nodes" in a social topology directly determine the rate of entropy increase in that region.

3. 模因学习与差别接触： 差别接触理论（Differential Association Theory）指出，犯罪是学习的结果 \cite{sutherland1992principles}。若一个个体接触到的“违法观念”（Pro-delinquent definitions）多于“反违法观念”，其决策算法将发生偏转。实验表明，社会学习理论（差别接触理论的社会心理学版本）对犯罪行为的解释力高达 68\% \cite{akers2017social}。这证明了社会拓扑中“邻居节点”的属性直接决定了该区域的熵增速度。

Long-term criminological practice has found that almost all criminals are linked to these three primary causes. While there are other risk-increasing triggers—such as economic interests, political goals, and psychological conflicts—these are merely the direct catalysts for potential criminals who already possess the primary causes. Individuals exposed to these triggers who lack the primary causes generally refrain from crime. However, if a person possesses two or more primary causes, they are highly likely to be a potential criminal and will likely commit an offense once a trigger occurs.

犯罪学经过长期的实践总结发现，几乎所有的犯罪分子都与这三个主因相关。虽然存在其他增加风险的诱因（如经济利益、政治目标、心理冲突），但诱因仅是促使已具备“主因”的潜在犯罪分子实施犯罪的直接催化剂。对于不具备这些主因的个体，即便面临风险诱因，通常也不会走向犯罪。然而，如果一个体同时具备两个或以上的主因，则其极大概率是潜在犯罪分子，一旦诱因出现，便极可能实施犯罪。

From a criminological perspective, if a society loses the hierarchical structure capable of maintaining the Rule of Law, it easily regresses into a "state of nature" or a jungle society. This is because memetic learning and differential association will fuel crime, causing a majority of the population to acquire at least two of the primary causes of crime. In such a scenario, traditional solutions—such as simply increasing the severity of punishment—will be insufficient to offset the systemic phase transition brought about by the surge in crime.

从犯罪学视角看，如果社会失去了足以维护法治的社会结构，那么该社会将不可避免地退回到“丛林社会”状态。这是因为模因学习与差别接触会助长犯罪扩散，导致社会中大多数节点最终都具备至少两个犯罪主因。在这种情况下，单纯依靠传统的犯罪处理方案（如盲目提高惩罚烈度），将难以对冲因犯罪激增而引发的系统相变。

\section*{Diverse Attributes of Enforcement Nodes and the Logic of Dissolution 执法节点的多元属性与消解逻辑}

In a system of the Rule of Law, "nodes" are not limited to domestic living individuals. Enforcement nodes can be symbolic entities that transcend time and space. Foreigners, the ideas of deceased luminaries, foreign legal experiences with universal value, or even protocols unrestricted by specific national boundaries can all play the role of "virtual nodes" within the social topology. These symbolic entities function as "memetic anchors" that provide standardized benchmarks for behavior across different epochs and geographies \cite{dawkins1976hierarchical, blackmore2000meme}.

法治体系中的“节点”并不局限于本国在世的个体。执法节点（Enforcement Nodes）可以是跨越时空的符号实体。外国人，已经去世的杰出人物的理念、具备普世价值的外国法治经验，甚至是不受特定国别限制的协议，都可以在社会拓扑中扮演“虚拟节点”的角色。这些符号实体作为“模因锚点”（Memetic anchors），为不同时代和地域的行为提供了标准化的参照基准 \cite{dawkins1976hierarchical, blackmore2000meme}。

An enforcement node must, at a minimum, consistently uphold a specific set of rules. For instance, Mao Zedong is regarded by some in China as a representative of the general public interest because his framework facilitated a distinct form of mass mobilization and collective assembly. If a node fails to maintain any coherent rules or "meta-protocols"—or if it merely promotes absolute egalitarianism or a Monolithic Leviathan—it struggles to function as an enforcement node dedicated to rule maintenance. Furthermore, an enforcement node need not be flawless; once its erroneous propositions are negated, the remaining framework can still function effectively as a robust enforcement node.

执法节点至少必须持续地维护一套特定的规则。例如，毛泽东在中国被部分人视为公众利益的代表，是因为其体制促成了一种形式的大众动员与集体集会。如果一个节点无法维持任何连贯的规则或“元协议（Meta-protocols）”，或者仅仅是推崇绝对平均主义或单一利维坦，那么它就很难履行其作为“维护规则者”的职能。此外，执法节点并不需要是完美的；在对其错误主张进行修正或否定后，其剩余的部分仍然可以作为优秀的执法节点继续运作。

According to the $J$ model established in this paper, these virtual nodes typically possess an extremely high fragility coefficient ($\beta \to 1$); they depend entirely on the universal recognition and procedural execution of society members. These "high-fragility nodes" serve as efficient levers to reduce the total cost of justice $J$ for the system. However, for a power center attempting to maintain a Monolithic Leviathan structure, these nodes are extremely dangerous competitors. This is because they provide "coordinates of justice" independent of the Leviathan \cite{habermas1975legitimation}. When other nodes are able to cite these virtual nodes as support for narrative legitimacy, their local coordination cost $\gamma$ decreases significantly, facilitating the emergence of spontaneous order \cite{hayek1973economic}.

根据本文的 $J$ 模型，这些虚拟节点通常具备极高的脆弱性系数（$\beta \to 1$），它们完全依赖于社会成员的普遍认同与程序化执行。这种“高脆弱性节点”是降低系统总执法成本 $J$ 的高效杠杆。然而，对于试图维持单一利维坦结构的权力中心而言，这些节点是极其危险的竞争者。因为他们提供了独立于利维坦之外的“正义坐标” \cite{habermas1975legitimation}。当其他节点能够引用这些虚拟节点作为叙事合法性支持时，其局部协调成本 $\gamma$ 会显著下降，从而促进自发秩序的涌现 \cite{hayek1973economic}。

Consequently, a Monolithic Leviathan state inevitably exhibits systemic "node erasure" behaviors. As Heinrich Heine famously warned, "Where they burn books, they will, in the end, burn people too" ("Dort, wo man Bücher verbrennt, verbrennt man auch am Ende Menschen") \cite{heine2021almansor}. From the perspective of our $J$ model, "burning books" represents the destruction of virtual nodes to monopolize interpretive power, while "burning people" represents the physical elimination of biological nodes that attempt to reactivate those rules.

因此，单一利维坦国家必然会表现出系统性的“节点抹除”行为。正如海涅（Heinrich Heine）那句著名的预言：“这只是序幕：在那烧书的地方，最后也将烧人。” ("Das war ein Vorspiel nur, dort wo man Bücher verbrennt, verbrennt man auch am Ende Menschen") \cite{heine2021almansor}。从 $J$ 模型的视角看，“烧书”是对虚拟节点的毁灭，旨在垄断解释权；而“烧人”则是对试图激活这些规则的生物节点的物理清除。

By plagiarizing and tampering with the achievements of advanced civilizations and packaging them as the Monolithic Leviathan’s "indigenous inventions," the state severs the connection between nodes and virtual nodes. The erasure of real historical heroes or successful cases of self-organization is, in essence, the removal of potential "historical enforcement nodes" within the topological structure that future generations could reference \cite{orwell2021nineteen}. By blocking external ideas to ensure a monopoly on interpretation within society, the state logically drives $\beta$ (power fragility) toward zero, thereby granting violators absolute immovability \cite{guriev2019informational}.

通过抄袭并篡改先进文明的成果，将其包装成单一利维坦的“内生发明”，从而切断节点与虚拟节点的连接。抹除历史上真实的英雄或成功自组织案例，其本质是移除拓扑结构中潜在的、可供后人引用的“历史执法节点” \cite{orwell2021nineteen}。通过封锁外部理念，确保社会内部只有单一的解释权，使 $\beta$（权力脆弱性）在逻辑上归零，从而让违规者获得绝对的不可动摇性 \cite{guriev2019informational}。

A Monolithic Leviathan typically prioritizes the suppression of low-fragility nodes—those characterized by high independence and low $\beta$ values. Such actions often elicit public acclaim, as they are superficially perceived as a liquidation of elite privilege, resonating with the primitive sense of justice rooted in absolute egalitarianism\cite{acemoglu2005economic}. However, this applause masks a profound systemic crisis. Low-fragility nodes function as the society’s "structural shock absorbers." When these nodes are systematically eliminated, the social architecture rapidly collapses from a stable Pyramid or Olive-shaped structure into an Hourglass form. Once this intermediate layer vanishes, high-fragility nodes at the bottom-tier are left directly exposed to the absolute will of the pinnacle. In the absence of mediating organizational structures, the local coordination cost $\gamma$ for the lower strata is infinitely magnified by algorithms and the Leviathan, resulting in the total atomization of individuals before state power.

单一利维坦还会优先打击低脆弱性节点（即那些具有较强独立性、较低 $\beta$ 值的节点）。这种行为往往容易赢得底层民众的拍手称快，因为从表面上看，这像是对精英特权的清算，符合“绝对平均主义”的朴素正义感\cite{acemoglu2005economic}。然而，这种掌声背后隐藏着严重的系统性危机。低脆弱性节点通常扮演着社会“减震器”的角色。当这些节点被系统性清除，社会结构便从稳固的金字塔型或橄榄型迅速坍缩为沙漏型。一旦中间层消失，底层的高脆弱性节点将直接面对顶层的绝对意志。此时，由于缺乏中间层的协调，底层的局部协调成本 $\gamma$ 会被算法和利维坦无限放大，导致个体在权力面前彻底原子化。

While the initial dismantling of low-fragility nodes may appear to reduce governance resistance, in the long term, the system loses its self-repairing and bottom-up oversight mechanisms. Consequently, the cost of maintaining rule of law, $J$, rises exponentially due to the failure of grassroots governance. In the transition toward a Rule-of-Law society, the treatment of intermediate nodes must be handled with extreme caution. A "top-to-bottom consistency" in rule enforcement is mandatory. Within the $J$ model, if rules are distorted during downward transmission, or if top-tier nodes are permitted to transcend these rules, a wholesale collapse of the intermediate layer ensues. Such a fracture inevitably plunges society into the Hourglass Trap, severing the buffer between the top and bottom and causing coordination costs $\gamma$ to skyrocket. 

虽然初期打击低脆弱性节点似乎降低了统治阻力，但从长期看，由于社会失去了自我修复和自下而上的监督机制，维护基本秩序的法治成本 $J$ 反而会因为基层治理的失效而呈指数级上升。在向法治社会转型的过程中，系统对于中间节点的处理必须极其谨慎。必须采取一种“自上而下一贯性”的规则执行标准。在 $J$ 模型中，如果规则在向下传递的过程中发生扭曲，或者顶层节点可以凌驾于规则之上，会导致中间层的整体坍缩。一旦中间层断裂，社会将不可避免地跌入“沙漏型陷阱”，使得顶层与底层之间失去缓冲，导致协调成本 $\gamma$ 激增。

Furthermore, rule consistency is vital for locking in a cooperative equilibrium within infinitely repeated games\cite{axelrod1981evolution}. Without it, enforcement nodes degenerate into mere "violence franchise holders." When society loses faith in the consistency of the protocols, the scope and horizon of the game contract rapidly. Even while an intermediate layer still exists physically, individuals are forced to seek optimal solutions within extremely narrow timeframes. This causes society to fall into total atomization at the game-theoretic level\cite{laibson1997golden}.

此外，规则一致性对于锁定长期多次重复博弈的合作均衡至关重要\cite{axelrod1981evolution}。若非如此，执法节点将沦为纯粹的“暴力特许经营者”。当社会成员不再信任规则的一致性时，博弈的广度（Scope and Horizon of the Game）会迅速缩减。即便中间层尚存，个体也只能在极短的时间尺度内寻找生存最优解，这使社会在博弈层面陷入彻底的原子化\cite{laibson1997golden}。

This evolutionary game failure is the micro-dynamic origin of the "Hourglass Trap" and the loss of control over coordination cost $\gamma$. Conversely, a Monolithic Leviathan deliberately employs selective enforcement in its early stages to create an atmosphere of terror. By artificially severing long-term expectations, it forces the game-theoretic atomization of society, systematically dismantling the maintenance structures of the rule of law. During this phase, the decisions of many social elites may appear "irrational" or "foolish"; however, these are in fact inevitable choices resulting from the drastic shrinkage of the game's scope\cite{laibson1997golden}.

这种演化博弈的失败，正是社会跌入“沙漏型陷阱”、协调成本 $\gamma$ 走向失控的微观动力学起源。反过来，单一利维坦在初期会刻意利用选择性执法，通过制造恐怖气氛人为切断博弈的长期性，迫使社会在博弈层面率先原子化，从而一步步肢解法治的维护结构。在这一阶段，许多社会精英的决策在观察者看来可能显得“愚蠢”或“短视”，但本质上，这是在博弈广度剧烈缩减、长期预期失效下的无可奈何之举\cite{laibson1997golden}。

\section*{Distributive Justice and Topological Stability 分配正义与拓扑稳定性}

Different distribution systems exert profound impacts on social topological structures. As discussed, the relationship between the total population $N$ and the hierarchical depth $L$ required to maintain the Rule of Law follows:

不同的分配制度会深刻影响社会拓扑结构。在上文中我们论述过，维护法治所需的系统的层级深度 $L$ 与 $N$ 的关系遵循：

$$N = k^L \implies L = \log_k N$$

The total population $N$ represents the sum of a geometric progression with an initial term of $1$ and a common ratio $k$:

而总人口是一个首项为 $1$，公比为 $k$ 的等比数列求和：

$$N = \sum_{i=0}^{L-1} k^i = \frac{k^L - 1}{k - 1}$$

The proportion of the population at the baseline tier, denoted as $P$, is given by:

而其基层比例为$P$：

$$P = \frac{N_{base}}{N} = \frac{k^{L-1}}{\frac{k^L - 1}{k - 1}} = \frac{k^{L-1}(k - 1)}{k^L - 1}$$

When the hierarchical depth $L$ is sufficiently large, $k^L \gg 1$, the formula can be approximated as:

当层级深度 $L$ 较大时，$k^L$ 远大于 $1$，公式可以近似为：

$$P \approx \frac{k^{L-1}(k - 1)}{k^L} = \frac{k-1}{k} = 1 - \frac{1}{k}$$

Given that $k$ is constrained by biological and cognitive limits such as Dunbar’s Number, let us assume $k=10$. The base-tier proportion then approximates:

$k$受到邓巴数等诸多限制，假设$k=10$，其基层比例为$$P \approx 1- \frac{1}{10}=\frac{9}{10}$$

Assume the per capita compensation at the base tier is $w$. Moving upward through the hierarchy, the per capita compensation follows a geometric progression:Tier $L$ (base): per capita $w$, node count $k^{L-1}$.Tier $L-1$: per capita $w(1+\lambda)$, node count $k^{L-2}$.Tier $L-i$: per capita $w(1+\lambda)^i$, node count $k^{L-1-i}$.Tier $1$ (Top): per capita $w(1+\lambda)^{L-1}$, node count $1$.

假设基层的人均报酬为 $w$。自下而上每一层的人均报酬呈等比级数增加：第 $L$ 层（基层）: 人均 $w$，节点数 $k^{L-1}$。第 $L-1$ 层: 人均 $w(1+\lambda)$，节点数 $k^{L-2}$。第 $L-i$ 层: 人均 $w(1+\lambda)^i$，节点数 $k^{L-1-i}$。第 $1$ 层（顶层）: 人均 $w(1+\lambda)^{L-1}$，节点数 $1$。

The total social income $W_{total}$ is the sum of (population $\times$ per capita compensation) for each tier:

总收入等于每一层的（人数 $\times$ 人均报酬）之和：$$W_{total} = \sum_{i=0}^{L-1} [k^{L-1-i} \cdot w(1+\lambda)^i]$$

Factoring out the constant $w \cdot k^{L-1}$:

提取常数 $w \cdot k^{L-1}$：

$$W_{total} = w \cdot k^{L-1} \sum_{i=0}^{L-1} \left( \frac{1+\lambda}{k} \right)^i$$

This represents a geometric series with a common ratio $q = \frac{1+\lambda}{k}$.

这是一个公比为 $q = \frac{1+\lambda}{k}$ 的等比数列。

The total wealth of the base tier is $W_{base} = k^{L-1} \cdot w$. The share of total wealth allocated to the base tier, $S_{base}$, is:

基层的总收入为 $W_{base} = k^{L-1} \cdot w$。基层分配比例 $S_{base}$ 为：$$S_{base} = \frac{W_{base}}{W_{total}} = \frac{1}{\sum_{i=0}^{L-1} (\frac{1+\lambda}{k})^i}$$

When $L$ is large and $\frac{1+\lambda}{k} < 1$, the denominator converges to the sum of an infinite geometric series $\frac{1}{1 - \frac{1+\lambda}{k}}$:

当 $L$ 较大且 $\frac{1+\lambda}{k} < 1$ 时，分母收敛于无穷等比级数 $\frac{1}{1 - \frac{1+\lambda}{k}}$：

$$S_{base} \approx 1 - \frac{1+\lambda}{k}$$

In empirical reality, the disparity in income across different hierarchical tiers is typically at least threefold. Assuming a factor of 3 (where $1+\lambda = 3$), and with $k=10$:

在现实中，不同层级的收入差异至少在三倍以上。假设就是3倍，当$k=10$的时候，$$S_{base} \approx 1 - \frac{1+2}{10}=\frac{7}{10}$$

The income share of the base tier primarily manifests as the labor share of GDP. Consequently, the labor share of GDP cannot exceed this theoretical ceiling; otherwise, it would undermine the structural integrity required to maintain the Rule of Law. This provides a structural explanation for why labor share of GDP, even in advanced Nordic social democracies, rarely exceed 70\%\cite{owid-labor-share-2020}. Attempts to push beyond this threshold frequently trigger a systemic collapse into a "Monolithic Leviathan" mode. Once the social structure essential for sustaining the Rule of Law is compromised, society inevitably descends into a state of mutual predation, where the ultimate stable equilibrium is the emergence of a Monolithic Leviathan \cite{Li_2026}.

这个基层总收入分配，主要就是劳动报酬占比。所以劳动报酬占比最高不能超过这个比例，否则会影响维护法治所需的结构。这也解释了为什么在现实中即使是实行民主社会主义的北欧发达国家，劳动报酬占比都没有超过70\%的\cite{owid-labor-share-2020}。而试图高于这个比例的尝试都很容易演化为单一利维坦模式。因为一旦失去维护法治所必须的社会结构，社会就会陷入互相掠夺的状态，最后稳定的解就是单一利维坦\cite{Li_2026}。

\section*{Conclusion: The Rule of Law as the Optimal Steady-State Solution for Complex Systems 结论：法治作为复杂系统的最优稳定解}

In previous research, we argued that the multilayered predatory stage is often fraught with Particularism, manifesting as identity-based differential pricing, asymmetric agreements, and "gangster logic" \cite{Li_2026}. Although such rules can maintain a transient equilibrium locally, their efficiency at a macro scale is extremely low. Lacking a unified philosophical or scientific fulcrum, the power of rule interpretation degenerates into a function of dynamic strength, causing the "Focal Point" in the Schelling sense to drift continuously \cite{schelling1980strategy}. Participants, unable to establish long-term stable expectations, are forced to consume scarce energy on defensive expenditures and contests over interpretive power. Particularistic rules are essentially high-entropy "ceasefire agreements"; lacking the anchorage of a Constitutional Commitment, any minute Fluctuation in power is amplified through positive feedback effects, inducing the system to cross a phase transition threshold and ultimately collapse into a Monolithic Leviathan mode.

在先前的研究中，我们曾论述：多层掠夺阶段往往充斥着特殊主义规则（Particularism），表现为身份差异化定价、非对称协议与黑帮逻辑 \cite{Li_2026}。尽管这种规则能在局部维持暂态平衡，但在宏观尺度上其效能极低。由于缺乏统一的哲学或科学支点，规则解释权沦为动态实力的函数，导致谢林（Schelling）意义上的“聚点”（Focal Point）处于持续漂移中 \cite{schelling1980strategy}。参与者因无法建立长期稳定的预期，不得不将稀缺能量消耗于防御性支出与解释权争夺。特殊主义规则本质上是一种高熵的“停火协议”，缺乏普世契约（Constitutional Commitment）的锚定，任何微小的力量涨落（Fluctuation）都会通过正反馈效应放大，诱发系统跨越相变阈值，最终坍缩为单一利维坦模式。

Combining the earlier derivation regarding the scale constraints of absolute egalitarianism and the quadratic growth of coordination costs ($O(k^2)$), we can conclude: in Large-scale Societies, to keep the total cost of rule maintenance within a threshold affordable by the system and to achieve high-resolution governance, a universal Rule of Law is the sole evolutionary optimal steady-state solution.

结合前文对绝对平均主义规模限制及协调成本平方级增长（$O(k^2)$）的推导，我们可以得出：在超大规模社会（Large-scale Societies）中，为使总体的规则维护成本处于系统可支付的阈值内并实现高分辨率治理，普遍性法治（Rule of Law）是唯一的演化最优稳定解。

In fact, within the long-term evolution of social topology, only three steady-state solutions exist:

事实上，在社会拓扑的长期演化中，仅存在三种稳态解：

1. Monolithic Leviathan: Achieves a low-dimensional "dead order" at the cost of sacrificing evolutionary kinetic energy by eliminating middle tiers.

1. 单一利维坦（Monolithic Leviathan）：通过消灭中间层，以牺牲演化动能为代价换取低维度的死寂秩序；

2. Small-scale Egalitarianism: Constrained by Dunbar’s number, it can only be sustained at a microscopic scale within a Fully Connected Graph.

2. 小规模绝对平均主义（Small-scale Egalitarianism）：受限于邓巴数，仅能在全连接图（Fully Connected Graph）的微型尺度下维持；

3. Hierarchical Society with Rule of Law: Achieves logarithmic cost compression through hierarchical depth $L$.

3. 多层级法治社会（Hierarchical Society with Rule of Law）：通过层级深度 $L$ 实现对数级的成本压缩。

Intermediate states other than these three (such as fragmented predation or crony particularism) are all metastable. Under evolutionary pressure, they typically collapse into one of the aforementioned three steady states within a few decades. Just as different species evolve similar traits under analogous ecological pressures—a phenomenon biologists call "Evolutionary Convergence"—diverse social structures, when subjected to the selection pressures of transaction costs and information entropy, inevitably converge toward these three topological "Attractors."

除此三者外的中间状态（如割据掠夺、裙带特殊主义等）均为亚稳态，在演化压力下通常会在数十年内坍缩至上述三种稳态之一。正如不同物种在相似的生态压力下会演化出相似的性状，生物学家称之为“演化的趋同性”，不同的社会结构在面对交易成本与信息熵的选择压力时，也必然向这三种拓扑“吸引子”（Attractors）收敛。

The reason the Rule of Law prevails at an ultra-large scale is that it provides a set of low-information-entropy "Meta-rules." These meta-rules are not a mere accumulation of empirical observations but are derived from philosophical principles or scientific laws. Consequently, their underlying logic is remarkably concise, possessing rigor and extreme universality. The goal of the Rule of Law is to terminate the state of "mutual predation." For nodes possessing a Rule of Law consciousness, a consistent standard for judging violations significantly lowers the entry threshold for coordination and counterbalancing. If a system suffers from immense controversy even in determining the degree of violation ($\epsilon$), coordination costs will once again fall into the trap of exponential growth.

法治之所以在超大规模下胜出，是因为它提供了一套低信息熵的“元规则”。这些元规则并非纯粹的经验堆砌，而是基于哲学原理或科学规律推导而出的，因此其底层逻辑十分简洁，具有严密性与极高的普适性。法治的目标是终结“互相掠夺”状态。对于具备法治意识的节点而言，一致的违规判定标准极大地降低了协调制衡的起点门槛。若系统在判定违规程度（$\epsilon$）上都存在巨大争议，协调成本将再次陷入指数级增长的陷阱。

The core challenge confronting contemporary society amidst the wave of automation, such as artificial intelligence and robotics, is the growing imbalance between the cost and benefit of maintaining functional hierarchical complexity. The social structure is evolving from the traditional olive-shaped or pyramidal form into an "Hourglass Society." This evolution not only portends the collapse of the middle class in economic terms but also creates an inherent technical dilemma for the maintenance of the Rule of Law.

当前社会在人工智能、机器人等自动化浪潮下面临的核心挑战是，维持功能性复杂层级的成本与收益比正在失衡。社会结构正在从传统的橄榄型或金字塔型演变为“沙漏型”。这种演变不仅在经济学上预示着中产阶级的坍缩，在技术上也造成了法治维护的内在困境。

According to the $J$ model established in this paper, the stability of the Rule of Law relies on a vast number of intermediate nodes characterized by high professionalism and high fragility (high $\beta$) to absorb coordination entropy and share the costs of maintaining justice. However, automation technology is essentially performing a "disintermediation" operation. Artificial intelligence is replacing a large number of technocrats and professional middle layers that were originally sustained by "Legal-rational Authority." This leads to a rupture in the $L$ dimension of the social topology, rendering the system possibly unable to achieve $O(\log N)$ cost compression through hierarchy. The impact of this predicament remains relatively mitigated in democratic nations. This is because democratic systems ensure that the composition of elected politicians maintains an olive-shaped or pyramidal structure—ranging from town mayors and city mayors to governors and heads of state, alongside their respective legislators—thereby enabling the maintenance of the fundamental Rule of Law. In contrast, in authoritarian states, the impact of this structural erosion is fatal. Furthermore, the pyramidal structure already harbors certain inherent risks, as the proportion of the population at the very base of society supporting absolute egalitarianism is considerably high. Consequently, economics most strongly advocates for the olive-shaped society. Practice has repeatedly demonstrated that only olive-shaped or pyramid-shaped societies can achieve common prosperity. This is because, in these structures, those at the pinnacle must drive the welfare growth of the entire society to achieve their own further development. Conversely, in an hourglass-shaped society, those at the top have effectively escaped the "gravitational pull" of the bottom-tier; the further development of the lower strata becomes entirely dependent on the benevolence or "alms" of the elite.

根据本文的 $J$ 模型，法治的稳定性依赖于大量具备高专业性、高脆弱性（高 $\beta$）的中间层节点来吸收协调熵并分担正义维护成本。然而，自动化技术实质上在执行“去中间化”操作。人工智能替代了大量原本由“法理型权威”支撑的技术官僚与专业中层。这导致了社会拓扑在 $L$ 维度上的断裂，使得系统可能无法通过层级实现 $O(\log N)$ 的成本压缩。这种困境在民主国家的影响还相对较小。因为民主制度决定了民选政客的组成肯定是一个金字塔型结构，从镇长到市长到州省长官到国家元首，还有相关的议员，从而可以维护基本的法治。但在威权国家，这种影响是致命的。而且金字塔型结构已经是有一定隐患了，因为社会最底层的民众支持绝对平均主义的比例相当高，所以经济学最推崇的是橄榄型社会。实践反复证明，只有橄榄型或金字塔型社会能实现共同富裕，因为深处塔尖的人必须要带动整个社会的福利增长，才能实现个人的进一步发展。而在沙漏型社会中，深处塔尖的人实际上脱离了底层的引力，底层的进一步发展完全取决于顶层的施舍。

At the dawn of the Internet, it was widely posited that the technology would significantly reduce coordination costs for the populace by eliminating physical barriers to communication \cite{shirky2008here}. However, in reality, these coordination costs have increased rather than decreased—or, at best, have remained largely unchanged. This paradox persists because, while the "transmission cost" of information has plummeted, the "processing cost" of information and the cost of reaching a consensus have instead surged due to "information overload" \cite{simon1996designing}. Furthermore, the Internet has fragmented the social landscape into a series of "Echo Chambers," creating formidable barriers to cross-group coordination \cite{sunstein2018republic}.

互联网诞生之初，有观点指出互联网可能降低民众的协调成本，因为它消除了沟通的物理障碍 \cite{shirky2008here}。但事实上，协调成本是变高了而不是变低了，最好的情况也是没什么变化。这种悖论之所以存在，是因为虽然信息的“传输成本”大幅下降，但由于“信息过载”，信息的“处理成本”和达成共识的成本反而激增了 \cite{simon1996designing}。此外，互联网导致社交景观碎片化为一个个“回声壁”（Echo Chambers），这为跨群体的协调制造了巨大的障碍 \cite{sunstein2018republic}。

From the perspective of the $J$ model proposed in this paper, the proliferation of high-entropy noise and the algorithmic manipulation by the "Digital Leviathan" have effectively weaponized information to ensure that individuals at the bottom-tier remain in an atomized state, thereby driving the local coordination cost $\gamma$ even higher \cite{tufekci2017twitter}.

从本文 $J$ 模型的视角看，高熵噪音的泛滥以及“数字利维坦”的算法操纵，实际上将信息武器化以确保底层民众保持原子化状态，从而使局部协调成本 $\gamma$ 进一步上升 \cite{tufekci2017twitter}。

An hourglass society consists of a tiny minority of super-nodes (low $\beta$) at the top and an immense, atomized populace at the bottom. The coordination costs required to counterbalance a node possessing data and computational hegemony grow geometrically, far exceeding the limits that the system can sustain. Under this structural configuration, the system easily crosses the phase transition threshold, collapsing into an Algorithmic Leviathan based on algorithmic governance\cite{konig2020dissecting,creel2022algorithmic}. Society should possess a comprehensive understanding and engage in robust discussions regarding this issue. By implementing a Universal Basic Income (UBI), we can reduce consumer price sensitivity, improve quality sensitivity, thereby steering social development toward refinement and differentiation and breaking free from the predicament of low-cost competition \cite{tuobang_2026_18267395}.

沙漏型社会由顶层的极少数超级节点（低 $\beta$）和底层极度原子化的庞大群体构成。制衡一个拥有海量数据与算力霸权的节点所需的协调成本呈几何级增长，远超系统所能负载的限度。在这种结构下，系统极易跨越相变阈值，坍缩为一种基于算法治理的单一利维坦（Algorithmic Leviathan）\cite{konig2020dissecting,creel2022algorithmic}。社会应该对此有充分的认识和讨论，通过发放全民基本收入降低消费者的价格敏感度，提升质量敏感度，从而将社会发展推向精细化，差异化，摆脱低成本竞争的困境\cite{tuobang_2026_18267395}。

\acknow{I acknowledges the Google Gemini in structuring the logic and refining the technical preparation of this work and the simulation online. I would also like to thank the support of peers from UC Berkeley during the preparation of this work.}

% --- 参考文献配置 ---
\bibliographystyle{unsrtnat} % 或者使用 pnas 风格：pnas-new
\begin{small}
    \bibliography{references} % 替换为你的 .bib 文件名，不需要加 .bib 后缀
\end{small}

% 结束中文环境
\clearpage
\end{CJK*}
\end{document}